\documentclass[apj,graphicx]{emulateapj}
\usepackage{apjfonts}

\newcommand\mxb{MXB~1659$-$29}
\newcommand\ks{KS~1731$-$260}
\newcommand\exo{EXO~0748$-$676}
\newcommand\xtej{XTE~J1701$-$462}
\newcommand\chand{{\it Chandra}}
\newcommand\xmm{{\it XMM-Newton}}

\begin{document}

\title{ Continued cooling of the crust in the neutron star low-mass X-ray binary
KS~1731$-$260}
\shortauthors{Cackett et al.}
\shorttitle{Crustal cooling of KS~1731$-$260}

\author{Edward~M.~Cackett\altaffilmark{1,2,6}, 
Edward~F.~Brown\altaffilmark{3},
Andrew~Cumming\altaffilmark{4}
Nathalie~Degenaar\altaffilmark{5},
Jon~M.~Miller\altaffilmark{1},
Rudy~Wijnands\altaffilmark{5}}

\email{ecackett@ast.cam.ac.uk}

\affil{\altaffilmark{1}Department of Astronomy, University of Michigan, 500
Church St, Ann Arbor, MI 48109-1042, USA}
\affil{\altaffilmark{2}Institute of Astronomy, University of Cambridge,
Madingley Rd, Cambridge, CB3 0HA, UK}
\affil{\altaffilmark{3} Department of Physics \& Astronomy, National
Superconducting Cyclotron Laboratory, and the Joint Institute for Nuclear
Astrophysics, Michigan State University, East Lansing, MI 48824, USA}
\affil{\altaffilmark{4}Department of Physics, McGill University, 3600 rue
University, Montreal QC, H3A 2T8, Canada}
\affil{\altaffilmark{5}Astronomical Institute `Anton Pannekoek', University of
Amsterdam, Kruislaan 403, 1098 SJ, Amsterdam, the Netherlands}

\altaffiltext{6}{{\it Chandra} Fellow}

\begin{abstract}
Some neutron star low-mass X-ray binaries (LMXBs) have very long outbursts
(lasting several years) which can generate a significant amount of heat in the
neutron star crust.  After the system has returned to quiescence, the crust then
thermally relaxes.  This provides a rare opportunity to study the thermal
properties of neutron star crusts, putting constraints on the thermal
conductivity and hence the structure and composition of the crust.  \ks{} is one
of only four systems where this crustal cooling has been observed.  Here, we
present a new \chand{} observation of this source approximately 8 years after
the end of the last outburst, and 4 years since the last observation.  We find
that the source has continued to cool, with the cooling curve displaying a
simple power-law decay.  This suggests that the crust has not fully thermally
relaxed yet, and may continue to cool further.  A simple power law decay is in
contrast to theoretical cooling models of the crust, which predict that the
crust should now have cooled to the same temperature as the neutron star core.
\end{abstract}


\section{Introduction}

Neutron stars in transient LMXBs provide a rare observational
opportunity to study the thermal properties of the neutron star crust.  In some
of these transient systems accretion outbursts last many years, as opposed to
the more typical weeks to months.  These quasi-persistent systems are
particularly interesting as during the long outbursts the crust should be heated
out of thermal equilibrium with the rest of the star \citep{rutledge_ks1731_02}.
 Thus, once the source returns to quiescence, the crust thermally relaxes.  The
resulting cooling curve depends on a number of key properties of the crust
including the composition and structure of the crust \citep{rutledge_ks1731_02,
shternin07, brown09}, and the crust thickness \citep[which is dependent on the
mass and radius of the star,][]{lattimer94, brown09}.
 
Such crustal cooling has now been observed in four neutron star transients:
\ks{} \citep{wijnands01, wijnandsetal02, cackett06}, \mxb{} \citep{wijnands03,
wijnandsetal04, cackett06, cackett08}, \exo{} \citep{degenaar09,degenaar10} and
\xtej{} \citep{fridriksson10}.  \ks{} was in outburst for 12.5 years, returning
to quiescence in 2001 \citep{wijnands01}.  For a detailed history of this
source, see \citet{cackett06}.  It was quickly realized that such a long
outburst should allow for observable crustal cooling, hence, a monitoring
campaign using \chand{} and \xmm{} followed
\citep{wijnands01,wijnandsetal02,cackett06}.  A cooling curve covering the first
 $\sim4$ years of quiescence was presented by \citet{cackett06}.  The source
flux initially dropped rapidly: a factor of $\sim$2.5 during the first year in
quiescence and a factor of $\sim5$ in the first 1000 days
\citep{wijnandsetal02,cackett06}.  The cooling curve was well fit by either an
exponential decay to a constant level or a simple power-law decay
\citep{cackett06,cackett08}.  The rapid rate of cooling and the inferred cold
temperature of the neutron star core, led to conclusions that the crust must
have a high thermal conductivity and required enhanced levels of core cooling
\citep{wijnandsetal02,cackett06} based on comparing with the theoretical cooling
models of \citet{rutledge_ks1731_02}.  Motivated by these cooling curves,
theoretical crust cooling models for \ks were calculated by
\citet{shternin07}.  These authors rule out a low thermal conductivity for the
crust, but do not require enhanced core cooling.  \citet{brown09} also
calculated thermal relaxation models for neutron star crusts, finding a
high thermal conductivity for the crust, and a low impurity parameter (i.e., the
dispersion in the charge of the ions in the crust is low).

Here, we present a new \chand{} observation of \ks, performed $\sim8$ years
after the end of the outburst, and 4 years since the previous \chand{}
observation.  The data are consistent with further cooling of the neutron star,
continuing along a power-law decay.

\section{Analysis and Results} \label{sec:results}

The newest \chand{} observation of \ks{} was performed in two separate
pointings.  One 31 ks segment was performed on 2009 May 17 (ObsID: 10037) and
the other 28 ks segment was performed on 2009 May 19 (ObsID: 10911).  The source
was at the default aimpoint on the ACIS-S3 chip, which was operated in FAINT
mode.  Given several recent changes to the calibration of \chand{}, we have
opted to also reanalyze all the previous \chand{} observations with the latest
calibration and software.  Moreover, we reanalyze the \xmm{} observations with
the latest calibration files and software. For full details of the previous
observations, please refer to \citet{cackett06}.  

\subsection{Chandra data reduction}

The \chand{} data were all analyzed using CIAO (v 4.2) and CALDB (v 4.2.2).  We
used a circular source extraction region with a radius of 3\arcsec, and the
background extraction region was an annulus with inner radius of 7\arcsec and
outer radius 25\arcsec.  The \verb|psextract| tool was used to extract the
spectra, and \verb|mkacisrmf| and \verb|mkarf| were used to create the response
matrices.

\subsection{XMM-Newton data reduction}

The \xmm{} data were analyzed using the XMM Science Analysis Software (v 9.0.0).
 The observation data files were reprocessed using the \verb|emproc| and
\verb|epproc| tasks. To check for high levels of background flaring we extracted
lightcurves for all events with $>10$ keV and pattern = 0 for the MOS, and 10 --
12 keV and pattern = 0 for the PN.  In all observations there was some
significant background flaring.  We excluded all times where the $>10$ keV
lightcurve was greater than 2 c/s for the MOS and 4 c/s for the PN.  We
extracted spectra using \verb|evselect| with a circular source extraction region
of radius 10\arcsec, and a circular background extraction region of radius
1\arcmin{} taken from a source free region close to the source.  For the MOS we
filtered with patterns 0 -- 12, and for the PN, we used  patterns 0 -- 4 and
flag = 0.  Response matrices were generated using \verb|rmfgen| and
\verb|arfgen|.

\subsection{Spectral analysis}

We fit the spectra using XSPEC \citep[ver. 12,][]{arnaud96}, following similar
procedures to \citet{cackett06} and \citet{cackett08}.  The spectra were modeled
with an absorbed neutron star atmosphere model.  We used the phabs model for
Galactic absorption, and the nsa model for the neutron star atmosphere
\citep{zavlinetal96}. Throughout, we fix the neutron star radius at 10 km, and
the neutron star mass at 1.4 M$_\odot$.  The normalization of the nsa model is
given by 1/$D^2$, where $D$ is the distance to the source in pc.  Here, we
assume a distance of 7 kpc \citep{muno00}, i.e. normalization =
$2.041\times10^{-8}$ pc$^{-2}$.  As we showed for MXB~1659$-$29 in
\citet{cackett08}, the distance assumed only shifts the fitted temperatures up
or down, and does not affect the cooling timescales derived.  Here, if we assume
a distance of 5 kpc, the temperatures are all about 10\% lower than for $D=7$
kpc, where as if we assume $D = 9$ kpc then the temperatures are approximately
7\% higher than for $D=7$ kpc.

We fit all the spectra simultaneously with the absorption column density,
$N_{\rm H}$, the same for all spectra, and a free parameter in the fit.  The
neutron star atmosphere effective temperature is allowed to vary between epochs.
 For the first \xmm{} observations (013795201/013795301) the two sets of spectra
had their parameters tied between the spectra.  Similarly, for the last \chand{}
observation (10037/10911) we also tie the parameters between the two spectra. 
Given the low count rates there are not enough counts per bin to use $\chi^2$
statistics when fitting, and therefore we use the W-statistic in XSPEC to fit
the unbinned spectra.

\begin{figure}
\centering
\includegraphics[angle=90,width=8cm]{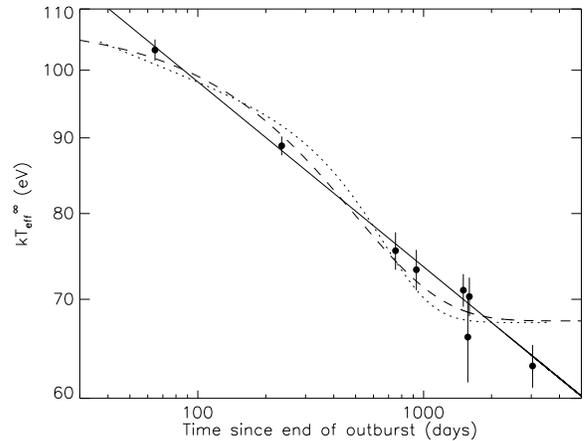}
\caption{Effective temperature for an observer at infinity for KS~1731$-$260
over approximately 3000 days from the end of the last outburst.  The solid line
shows a power-law fit to these temperatures, while the dashed line shows the
best-fitting exponential decay to a constant level.  The dotted line shows the
best-fitting model using the crustal cooling simulations from \citet{brown09}.}
\label{fig:coolingcurve}
\end{figure}

The results of the spectral fitting are given in Table~\ref{tab:specfits}, and
the evolution of the effective temperature is shown in
Figure~\ref{fig:coolingcurve}.  All uncertainties quoted and plotted are at the
1$\sigma$ level of confidence.  Note that in \citet{cackett06} the uncertainties
quoted were at the 90\% level of confidence, not 1$\sigma$ as stated in the
text.  The results clearly show a decrease in the neutron star atmosphere
temperature over time, with the newest observation the coldest yet.

\begin{deluxetable*}{lcccccccc}
\tabletypesize{\scriptsize}
\tablecolumns{9}
\tablewidth{0pc}
\tablecaption{Neutron star atmosphere spectral fitting parameters}
\tablecomments{All uncertainties are 1$\sigma$. The nsa atmosphere model is used
and the mass and radius are fixed in all fits to 1.4 M$_\odot$ and 10 km
respectively.  $kT_{\rm eff}^{\infty}$ is the effective temperature for an
observer at infinity, i.e. the gravitationally redshifted effective temperature.
 The normalization of the atmosphere was set for a distance = 7 kpc.  ObsIDs of
each observation are indicated at the top, with CXO (\chand{}) and XMM (\xmm{})
denoting the observatory.}
\tablehead{  & 2428 & 013795201/301 & 3796 & 3797 & 0202680101 & 6279 & 5468 &
10037/10911 \\
 & (CXO) & (XMM) & (CXO) & (CXO) & (XMM) & (CXO) & (CXO) & (CXO) }
\startdata
MJD & 51995.1 & 52165.7 & 52681.6 & 52859.5 & 53430.5 & 53500.4 & 53525.4 &
54969.7\\
N$_{\rm H}$ ($10^{22}$ cm$^{-2}$) & \multicolumn{7}{c}{$1.30\pm0.06$} \\
$kT_{\rm eff}^{\infty}$ (eV) & $103.2\pm1.7$ & $88.9\pm1.3$ & $75.5\pm2.2$ &
$73.3\pm2.3$ & $71.0\pm1.8$ & $66.0\pm4.5$ & $70.3\pm2.1$ & $63.1\pm2.1$\\ 
$F_{\rm bol}$\tablenotemark{a} (10$^{-13}$ erg cm$^{-2}$ s$^{-1}$) & $4.2\pm0.3$
& $2.3\pm0.1$ & $1.2\pm0.1$ & $1.1\pm0.1$ & $0.95\pm0.09$ & $0.72\pm0.21$ &
$0.92\pm0.11$ & $0.60\pm0.08$
\enddata
\label{tab:specfits}
\tablenotetext{a}{Bolometric flux calculated from the model over the 0.01 - 100
keV range}
\end{deluxetable*}

\subsection{Count rate analysis}

One potential problem with our interpretation of crustal cooling is that
there could be a power-law spectral component that cannot be detected here
because of the low number of counts in the spectra.  Non-thermal power-law
components are common in many quiescent neutron star spectra
\citep[e.g.][]{jonker04}.  Any power-law component could potentially change the
shape of the cooling curve of the thermal component.  It should be noted that
the highest quality spectra we have of \ks{} is the first \chand{} observation. 
This observation was discussed in detail by \citet{wijnands01}.  These authors
show that the spectrum is consistent with a thermal spectrum only, and a
power-law is not required statistically.  However, if they add a power-law to
their model they find that it only contributes $\sim$15\% to the flux.  Due to
the low count rate of the later observations, here, we use count ratios between
different bands to assess whether there could be a significant power-law
component present.  One can only do this with observations from the same
telescope (\chand{} and \xmm{} have different effective areas), thus we analyze
just the raw counts from the six \chand{} observations.

\begin{figure}
\centering
\includegraphics[angle=90,width=8cm]{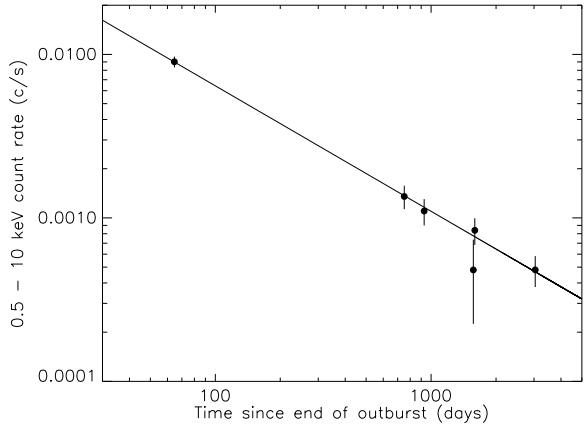}
\caption{0.5-10 keV count rates for all {\it Chandra} observations of
KS~1731$-$260.  The solid line shows a power-law fit to these count rates.}
\label{fig:rate}
\end{figure}

\begin{figure}
\centering
\includegraphics[angle=90,width=8cm]{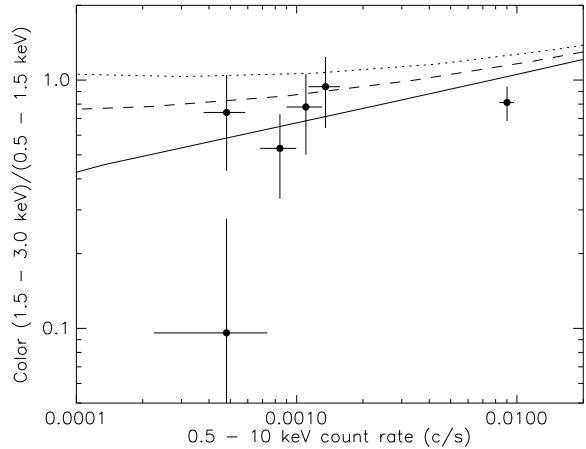}
\caption{Hardness-intensity diagram for {\it Chandra} observations of
KS~1731$-$260.  The hardness color is the ratio of the 1.5 --3.0 keV count rate
to the 0.5 -- 1.5 keV count rate.  This is plotted against the count rate for
the 0.5 -- 10 keV band.  The solid line shows the predicted count rates for a
cooling neutron star atmosphere.  The dashed line shows the predicted count
rates for a neutron star atmoshere and power-law spectrum, where the power-law
($\Gamma = 2$) always contributes 20\% to the unabsorbed 0.5 -- 10 keV flux. 
The dotted line shows a neutron star atmosphere plus power-law spectrum ($\Gamma
= 2$) where the power-law flux always contributes 40\% to the unabsorbed 0.5 --
10 keV flux.}
\label{fig:hid}
\end{figure}

Figure~\ref{fig:rate} shows \chand{} count rate versus time.  This lightcurve
also shows a power-law decay like the temperatures from spectral fitting. 
We find that only the very first observation has a significant detection in the
3 -- 10 keV band where the power-law component would dominate.  To study the
spectral evolution of the source, we therefore define a hardness color ratio
between the count rate in the 1.5 -- 3.0 keV band and the 0.5 -- 1.5 keV band. 
In Figure~\ref{fig:hid} we show a hardness-intensity diagram that displays the
evolution of this color with the full 0.5 -- 10 keV count rate.  We also
show the predicted color and count rate evolution for three different spectral
models: (i) a neutron star atmosphere, (ii) a neutron star atmosphere plus
power-law model where the power-law always contributes 20\% to the unabsorbed
0.5 -- 10 keV flux (we assume a spectral index, $\Gamma = 2$), (iii) a neutron
star atmosphere plus power-law ($\Gamma = 2$) model where the power-law flux
always contributes 40\% of the unabsorbed 0.5 -- 10 keV flux (note that all
three models assume Galactic absorption of $N_{\rm H} = 1.3\times10^{22}$
cm$^{-2}$).   The data points are most consistent with a simple cooling neutron
star atmosphere, but, model (ii) where there is a small contribution from a
power-law cannot be ruled out.  Any power-law contribution increases the color
(for a given count rate) compared to the color from a neutron star atmosphere
only.   A harder power-law spectrum (with a slope closer to 1 than 2) would also
lead to an increased color. Thus, any contributions from a power-law component
must be at a low level (less than a few tens of percent).  Therefore, there must
still be significant neutron star cooling.  Also note that this count rate
analysis does not take into account any change in the effective area of the
detector over time.  There has been a known increase in contaminant on the
\chand{} ACIS detector since the mission launch, leading to a decrease in the
sensitivity at lowest energies over time \citep{marshall04}. This would
artificially harden the color used here, thus the colors shown here should be
taken as maximum values. 

To further test the affect of a constant fraction of power-law flux, we fit the
spectra with an absorbed neutron star atmophere plus power-law model.  We fix
the power-law with slope $\Gamma = 2$ and the normalization for each observation
is set to give a 0.5 -- 10 keV unabsorbed flux that is 20\% of the value found
from fitting a neutron star atmosphere alone.  The resulting temperatures still
follow a simple power-law decay cooling curve, with the same slope as before. 
This can be easily understood, as reducing the thermal flux by 20\% will simply
just reduce the measured temperature by approximately 5\% as $F\propto T^4$.

\subsection{Cooling curves}

The latest \chand{} observation analyzed here was taken approximately 4 years
after the nearest observations, and appears to show continued cooling of
the neutron star.  We compare the three observations taken around MJD 53500 with
this newest observation.  These three observations (the fifth, sixth and seventh
quiescent observations) were all taken within 100 days of each other,
and all have consistent temperatures at the 1$\sigma$ level.  A weighted average
of the temperatures from these three observations gives $kT_{\rm eff}^{\infty} =
70.3\pm1.3$ eV.  Compared to the latest \chand{} observation, where $kT_{\rm
eff}^{\infty} = 63.1\pm2.1$ eV, the two temperatures differ at the 3$\sigma$
level of confidence, suggesting continued cooling of \ks.

Before this latest observation, the cooling curve could be well fit by either a
power-law \citep{cackett08} or an exponential decay to a constant level
\citep{cackett06}.  We tested both forms here.  We fitted a power-law of the
form $y(t) = \alpha(t-t_0)^\beta$ to the effective temperatures\footnote{Note
that as $F\propto T^4$ the flux decay curve can simply be calculated from the
fit to the temperatures}, where we chose $t_0$ to be midday on the last day the
source was observed to be active \citep[MJD 51930.5, see][]{cackett06}.  The
best fitting parameters are $\alpha=174.7\pm1.3$ eV and $\beta=-0.125\pm0.007$,
giving $\chi^2_\nu = 0.33$, $P_\chi = 0.92$.  This best-fitting power-law is
shown in Figure~\ref{fig:coolingcurve} as a solid line, and fits the cooling
curve well.  For comparison, we also test an exponential decay of the form $y(t)
= a \exp[-(t-t_0)/b]  + c$.  The best fitting parameters are $a = 39.8\pm2.3$
eV, $b = 418\pm70$ days and $c = 67.7\pm1.3$ eV, giving $\chi^2_\nu = 2.0$,
$P_\chi = 0.04$.  The simple power-law decay is therefore a better fit to the
cooling curve, further indicating continued cooling of \ks.

We also fitted numerical crust cooling simulations from \citet{brown09} to
the data. These calculations by \citet{brown09} suggest that the
lightcurve of a cooling crust is expected to be a broken power-law which
flattens to a constant at late times, set by the temperature of the neutron star
core.  The initial power-law decay is set by the temperature profile in the
outer crust.  The power-law break occurs when there is a transition in the solid
from a classical to quantum crystal at a depth close to neutron drip.  The time
of this power-law break is set by the thermal diffusion time to the depth of
this transition.  In these models the three variable parameters are (i) the
crust impurity parameter which is given by $Q_{\rm imp} \equiv n^{-1}_{\rm ion}
\displaystyle\sum_{i} n_i (Z_i - \langle Z \rangle)^2$   and measures the
dispersion in the charge of the nuclides ($Z$) in the crust, (ii) the core
temperature, $T_c$ and (iii) the temperature at the top of the crust, $T_b$. 
The best-fitting model (dotted line in Fig~\ref{fig:coolingcurve}) has $Q_{\rm
imp} = 4.0$, $T_c = 5.4\times10^7$ K, and $T_b = 2\times10^8$ K (at a column of
$1\times10^{12}$~g~cm$^{-2}$ during the outburst).  This model gives $\chi^2_\nu
= 3.3$.  These values are quite similar to the best-fit in \citet{brown09} who
found $Q_{\rm imp} = 1.5$, $T_c = 4.6\times10^7$ K, and $T_b = 2.5\times10^8$ K.
 Thus, the core temperature here is slightly colder and the impurity parameter
in the crust is slightly higher than the previous best-fitting values.

\section{Discussion and Conclusions} \label{sec:disc}

We have presented a new \chand{} observation of the neutron star low-mas X-ray
binary \ks{} in quiescence.  This observation extends the monitoring of this
source to approximately 8 years after the end of the most recent outburst.  The
observation suggests that the neutron star has continued to cool with an
effective temperature (for an observer at infinity) of $kT_{\rm eff}^{\infty} =
63.1\pm2.1$ eV compared to the preceding epoch where $kT_{\rm eff}^{\infty} =
70.3\pm1.3$ eV.  Previously, the cooling curve could be well fit by both a
power-law decay or an exponential decay to a constant level
\citep{cackett06,cackett08}.   Here, we find that the latest observation remains
on the simple power-law decay with a slope of $-0.125\pm0.007$, indicating that
the crust may be continuing to cool and has not yet reached thermal equilibrium
with the core.  As has been discussed previously
\citep{wijnandsetal02,cackett06,shternin07,brown09}, the cooling of \ks{}
indicates a crust with high thermal conductivity.  This implies that the crust
has a low impurity parameter, in other words, the dispersion in the charge of
the ions in the crust is low \citep[][find $Q_{\rm imp} < 10$ is a robust upper
limit]{brown09}.  

One potential problem is that the spectrum may not be purely thermal, but that
there could also be some contribution from non-thermal power-law emission that
cannot be detected due to low number of counts in the spectra.  Recent
observations of another cooling neutron star \xtej{} have shown an increase in
flux for approximately 100 days on top of a cooling curve, likely due to
increased levels of residual accretion \citep{fridriksson10}.  Furthermore,
long-term monitoring of the neutron star transient Cen~X-4 in quiescence has
shown variability in the thermal component that cannot be explained by crustal
cooling, and may also be linked to residual accretion during quiescence
\citep{cackett10}.  The monitoring of \ks{} is relatively sparse (only 8
observations over 8 years) and thus any increase in flux similar to \xtej{}
could easily have been missed.  However, only the first observation of \ks{}
shows a significant detection above 3 keV, and in this observation
\citet{wijnands01} found that the contribution of the power-law was limited to
about 15\% of the flux, though the power-law component was not statistically
required to fit the spectrum.  In order to investigate the potential
contribution from power-law emission, we created a hardness-intensity diagram
using only the \chand{} count rates in different bands.  The data are consistent
with a cooling neutron star atmosphere, though a small contribution (a few tens
of percent) from a power-law component cannot be excluded.

From a theoretical perspective, it is puzzling that the decay is best-fit by a
single power law.  Numerical simulations of crustal cooling expect that the
lightcurve should approximate a broken power-law which flattens to a constant at
late times \citep{brown09}.  The break in the power-law is caused by a change
from classical to quantum crystals, and the timescale on which this occurs is
set by the thermal diffusion time to the depth of this transition.  Though in
\ks{} we only see a single, unbroken power-law decay in the lightcurve, fitting
the model of \citet{brown09} allows for quantitative constraints on the thermal
conductivity of the crust.   The best-fitting model gives an impurity parameter
$Q_{\rm imp} = 4.0$ and core temperature $T_c = 5.4\times10^7$ K  (note that
these values are for a fixed mass and radius, and any increase in the surface
gravity shortens the cooling time by decreasing the crust thickness).  The
impurity parameter is marginally higher and the core temperature is slightly
lower than the best-fitting values in \citet{brown09}.  The additional \chand{}
observation requires a lower core temperature to fit the lightcurve at late
times, where as the impurity parameter needs to be increased to maintain the
thermal cooling timescale (a lower core temperature increases the thermal
conductivity and thus increasing the impurity parameter compensates).   The
implied core temperature only requires standard models for core cooling without
enhanced levels on neutrino cooling \citep{shternin07,brown09}.

It is interesting to note that a power-law decay with a slope of 1/8 (as
observed here) is expected for thermal relaxation of the core by enhanced
neutrino emission \citep[i.e., the rate goes as $T^6$, see e.g.,][]{page06}.
Such a scenario requires the core to be heated up substantially during an
outburst. However, in order to do this, or to cool the core over the observed
timescale, would require a core specific heat several orders of magnitude
smaller than that provided by degenerate electrons, assuming an electron
fraction of order 0.1.

\acknowledgements
\vspace{-0.25cm}
EMC thanks the staff at Durham University for kind hospitality during a visit
where much of this work was completed. EMC gratefully acknowledges support
provided by NASA through the {\it Chandra} Fellowship Program.

\bibliographystyle{apj}

\end{document}